\documentclass[aps,prx,superscriptaddress,showpacs,twocolumn]{revtex4-2}
\usepackage[left=1.7cm,right=1.7cm,bottom=1.7cm,top=1.7cm,ignoreall]{geometry}
\usepackage[utf8]{inputenc}
\usepackage[english]{babel}
\usepackage{csquotes}
\MakeOuterQuote{"}
\usepackage{graphicx}
\usepackage{amsmath}
\usepackage{amssymb}
\usepackage{xspace}
\usepackage{bm}
\usepackage{color}
\usepackage[squaren,Gray]{SIunits}
\usepackage[hyperindex=true]{hyperref}
\hypersetup{linktocpage,colorlinks=true,citecolor=blue,linkcolor=blue}
\usepackage{empheq}
\usepackage{braket}
\usepackage{physics}
\DeclareMathOperator{\sinc}{sinc}

\begin{document}
	
\title{Tailoring quantum walks in integrated photonic lattices}

\author{A.\ Raymond}
\affiliation{Université Paris Cité, CNRS, Laboratoire Matériaux et Phénomènes Quantiques, 75013 Paris, France}

\author{P.\ Cathala}
\affiliation{Université Paris Cité, CNRS, Laboratoire Matériaux et Phénomènes Quantiques, 75013 Paris, France}

\author{M.~Morassi}
\affiliation{Université Paris-Saclay, CNRS, Centre de Nanosciences et de Nanotechnologies, 91120 Palaiseau, France}

\author{A.~Lemaître}
\affiliation{Université Paris-Saclay, CNRS, Centre de Nanosciences et de Nanotechnologies, 91120 Palaiseau, France}

\author{F.~Raineri}
\affiliation{Université Côte d’Azur, Institut de Physique de Nice, CNRS-UMR 7010, 06200 Nice, France}
\affiliation{Université Paris-Saclay, CNRS, Centre de Nanosciences et de Nanotechnologies, 91120 Palaiseau, France}
	
\author{S.~Ducci}
\affiliation{Université Paris Cité, CNRS, Laboratoire Matériaux et Phénomènes Quantiques, 75013 Paris, France}

\author{F.~Baboux}
\thanks{\textcolor{blue}{florent.baboux@u-paris.fr}}
\affiliation{Université Paris Cité, CNRS, Laboratoire Matériaux et Phénomènes Quantiques, 75013 Paris, France}
	
\makeatletter
\def\Dated@name{} 
\makeatother

\date{}

	
\begin{abstract}

Unlike discrete photonic circuits, which manipulate photons step-by-step using a series of optical elements, arrays of coupled waveguides enable photons to interfere continuously across the entire structure. When composed of a nonlinear material, such arrays can also directly generate quantum states of light within the circuit. To clarify the similarities and distinctions between these two approaches of quantum walks, we conduct here a systematic comparison between linear waveguide arrays, injected with photons produced externally, and nonlinear arrays, where photon pairs are continuously generated via parametric down-conversion. We experimentally validate these predictions using III-V semiconductor nonlinear waveguide lattices with varied geometries, enabling us to tune the depth of the quantum walks over an order of magnitude and reveal the gradual emergence of non-classicality in the output state. Finally, we demonstrate an inverse-design approach to engineer \textit{aperiodic} waveguide arrays, whose optimized coupling profiles generate maximally entangled states such as the biphoton W-state. These results highlight the potential of continuously-coupled photonic systems to harness high-dimensional entanglement within compact architectures.

\end{abstract}
	
\maketitle

\section{Introduction}

Quantum photonic circuits have emerged as a promising platform for implementing quantum information tasks, from communication to computation and sensing applications \cite{Wang20}. A natural approach to encoding quantum states in these systems is through path encoding, where information is carried by waveguide modes~\cite{Solntsev17}. This method inherently supports high-dimensional quantum states, providing increased computational power and resilience to noise compared to two-level systems \cite{Erhard20}. In this context, two distinct paradigms have emerged. 
The first approach uses multicomponent photonic circuits where path-encoded quantum states are manipulated sequentially via discrete optical elements such as quantum state sources, beamsplitters, and phase shifters. While much of the footprint is dedicated to routing waveguides, this method efficiently implements a gate-based quantum computing model \cite{Knill01} and enables discrete-time quantum walks~\cite{Aharonov93} for tasks like on-chip Boson Sampling \cite{Brod19}. An alternative approach leverages instead continuously-coupled waveguides, where photons spread transversely, interfering along the entire propagation length rather than solely at discrete beamsplitters~\cite{Christodoulides03,Hoch22}. This enables a high connectivity within a compact footprint and naturally implements continuous-time quantum walks \cite{Perets08}, linking photonic dynamics to lattice Hamiltonians and condensed matter physics \cite{Farhi98}.
Interestingly, even greater possibilities arise when the waveguide array is composed of a nonlinear material, enabling the direct generation of quantum states within the array through parametric processes such as spontaneous parametric down-conversion (SPDC) \cite{Solntsev12,Kruse13,Solntsev14,Raymond24} or four-wave mixing (SFWM) \cite{BlancoRedondo18,Wang19,Doyle22,Ren22} . This results in a system where photon generation and continuous interference occur simultaneously, leading to novel dynamics that have no direct counterpart in bulk optics or discrete photonic circuits \cite{Hamilton14,Antonosyan14,Luo19,Belsley20,Barral20,Barral21,Hamilton22,He24,Baboux24,Costas25,Delgado25,Delgado25b}.

While both linear and nonlinear waveguide arrays have demonstrated remarkable capabilities for quantum photonics, they have largely been studied independently up to now. A comprehensive understanding of their similarities and distinctions would be instrumental in designing hybrid approaches that leverage the advantages of each.
To this end, we provide here a systematic comparison between quantum walks in linear and nonlinear waveguide arrays, combining analytical and numerical approaches. 
We first show, through different examples, how the output state of a nonlinear array can be understood as a coherent superposition of quantum walks in linear arrays of continuously varying lengths—an interpretation that provides valuable intuition into the underlying system dynamics. 
We then validate these predictions with experiments carried out in AlGaAs semiconductor nonlinear waveguide arrays with varying geometrical parameters, allowing to tune the depth of the quantum walks over an order of magnitude. We analyze the stabilization of the correlation pattern and the gradual emergence of non-classicality using two distinct criteria, revealing different aspects of path entanglement.
Finally, we present an inverse-design approach to engineer aperiodic waveguide lattices with spatially inhomogeneous coupling constants, demonstrating that it enables the generation of maximally entangled states from moderate experimental resources. We also outline possible hybrid schemes combining linear and nonlinear arrays on a single chip, which could expand the possibilities for quantum information tasks such as quantum state engineering or quantum state characterization.

\begin{figure*}[t]
	\centering
	\includegraphics[width=14cm]{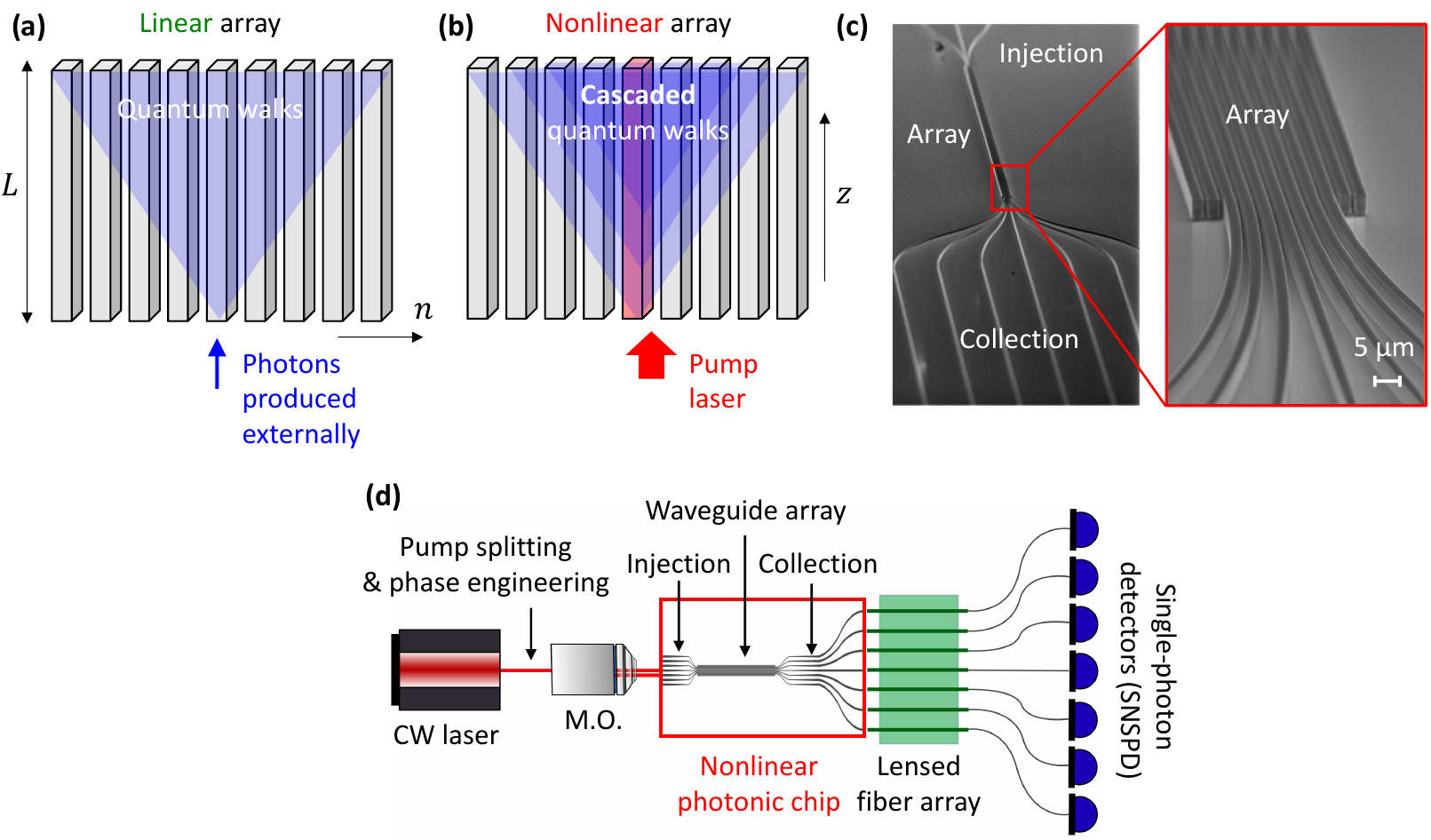}
	\caption{
		(a) Schematic representation of quantum walks (sketched in blue) in a linear waveguide array injected with externally produced photons and (b) cascaded quantum walks of photons generated internally by parametric down-conversion in a nonlinear waveguide array.
		(c) SEM images of an AlGaAs nonlinear waveguide array surrounded by S-bent waveguides for the injection and collection of light in and from the different waveguides.
		(d) Sketch of the experimental setup (M.O. = microscope objective, SNSPD = superconducting nanowire single-photon detectors).
	}
	\label{Fig_Setup}
\end{figure*}

\section{Single-waveguide input}

\subsection{Linear waveguide array}

To analyze the effect of continuous photon interference in waveguide arrays, we first consider the elementary building block consisting of two coupled waveguides, injected with two indistinguishable single photons.
For two identical waveguides of length $L$ and coupling constant $C$, if one photon is injected in each waveguide, the probability of bunching (both photons exiting from the same waveguide) is $\sin^{2}{(2CL)}$, while the probability of antibunching (photons exiting from separate waveguides) is $\cos^{2}{(2CL)}$. Hence, perfect bunching (resp. antibunching) is obtained for lengths $L = \left( m + \frac{1}{2} \right)L_{c}$ (resp.\ $ L= mL_{c}$), where $L_{c} = \frac{\pi}{2C}$ is the half-beat length and $m$ is an integer.

By contrast, in a waveguide array the continuous spreading of the wavefunction over a high number of waveguides significantly alters this behavior of periodic alternation of bunching and antibunching.
To illustrate this, let us consider a periodic array of identical and equidistant waveguides labeled by an integer $n$, with two indistinguishable photons injected in the central waveguide $n \!=\! 0$ (Fig.~\ref{Fig_Setup}a), i.e.~in the separable state $\ket{00}$.
Each photon propagates transversely in the array, giving rise to a so-called discrete diffraction pattern~\cite{Christodoulides03}. Assuming an infinite array, the biphoton wavefunction at the output (position $z=L$), i.e.~the probability amplitude to detect one photon in waveguide labeled $n_s$ and the other photon in waveguide $n_i$, reads
\begin{equation}
	\phi_{n = 0}\left( n_{s},n_{i}, L \right) = i^{n_{s} + n_{i}}\ J_{n_{s}}(2CL)\ J_{n_{i}}(2CL)
	\label{phi0}
\end{equation}
where \(J_{n}\) is the first-kind Bessel function of order $n$. This corresponds to a separable state. 
The calculated intensity correlation map $\left| \phi_{n = 0} \left( n_{s},n_{i} \right) \right|^{2}$ is shown in Figs.~\ref{Fig_Passive}a-b-c for increasing values of the normalized propagation length $CL$ -- which controls the depth of the quantum walks --  from $CL=0.3$ to $CL=5$.
As expected from the separable character of the state, these correlation maps do not favor bunching nor antibunching; after a certain propagation length, they fill a square pattern, where the four vertices correspond to a ballistic propagation of the two photons (characteristic of a disorder-free system) \cite{Lahini08}, reaching a distance of order $\pm 2CL$ (in waveguide number) on each side of the central waveguide \cite{Bromberg09}. 
At small propagation lengths (Fig.~\ref{Fig_Passive}a), the off-diagonal terms (e.g.\ $\phi_{n = 0} (0,1) \propto J_0 J_1$) are still larger (in modulus) than the "ballistic" terms (e.g.\ $\phi_{n = 0} (1,1)$ and $\phi_{n = 0} (1,-1)$, which are $ \propto J_1^2$); these terms equalize when $ J_1= J_0$, which first occurs for $CL\simeq0.72$. After this, the "ballistic" pattern with dominant diagonal and antidiagonal lobes stabilizes, and the correlation map essentially evolves as a dilatation, with the positions of the ballistic lobes increasing linearly with \(L\) (Figs.~\ref{Fig_Passive}b and c).
The wavefunction spreads over more waveguides but the Schmidt number, which quantifies the effective number of modes needed to decompose the biphoton wavefunction, remains pinned to 1 as $CL$ increases, as shown in Fig.~\ref{Fig_Passive}e (green line), since propagation in a passive optical circuit cannot increase entanglement \cite{Sperling11} (although it can transform one form of entanglement into another \cite{LoFranco18}).

\subsection{Nonlinear waveguide array}

Let us now consider a nonlinear waveguide array, where photon pairs can be generated continuously through SPDC by injecting a pump laser beam (whose transverse coupling can be neglected due to its twice shorter wavelength) into the central waveguide \cite{Solntsev12}, as sketched in Fig.~\ref{Fig_Setup}b. 
Since the photon pairs can be generated with equal probability at all possible longitudinal positions $z$, from an intuitive reasoning we expect the output state to correspond to the superposition of quantum walks started at different $z$.
This can be expressed analytically by starting from the biphoton state in momentum space, for which a closed-form expression can be obtained.
In the case of a monochromatic pump beam, the spatio-spectral wavefunction at the output of the nonlinear waveguide array reads \cite{Solntsev12,SM}:
\begin{equation}
	\widetilde{\Psi}(\omega_s,\omega_i,k_s,k_i) = \gamma L \, \widetilde{A}_p(k_s+k_i) \, \sinc \left( \Delta\beta \,L/2 \right)  e^{i \Delta\beta \,L/2}
\end{equation}
where $\omega_s$ and $\omega_i$ are the (angular) frequencies of the signal and idler photons, $k_{s}$ and $k_{i}$ are their dimensionless transverse wavevectors (normalized to $1/a$, where $a$ is the lattice periodicity), $\gamma$ quantifies the efficiency of the SPDC process, $\widetilde{A}_p$ is the spatial Fourier transform of the pump beam (determined by the amplitude of the pump field in each waveguide), and $ \Delta\beta = \beta_p -  \beta_s - \beta_i  $  is the phase-mismatch between the pump ($p$), signal ($s$) and idler ($i$) modes. For the latter two modes, the evanescent coupling at telecom wavelength gives rise to a Bloch-like dispersion relation for the propagation constants:
\begin{equation}
	\beta_{s,i}(\omega_{s,i}, k_{s,i})=\beta_{s,i}^{(0)}(\omega_{s,i})+2\,C_{s,i} (\omega_{s,i})\,\text{cos} (k_{s,i})
	\label{eq:band}
\end{equation}
where $\beta_{s,i}^{(0)}$ is the propagation constant of a single (uncoupled) waveguide and $C_{s,i}$ are the evanescent coupling constants between waveguides. Neglecting the transverse coupling of the pump beam, $\Delta\beta$ can be written as
\begin{multline}
	\Delta\beta(\omega_s, \omega_i, k_s, k_i)=\Delta\beta^{(0)}(\omega_s, \omega_i)\\
	-2\,C_{s}(\omega_s) \,\text{cos}(k_s)-2\,C_i(\omega_i) \,\text{cos}(k_i)
	\label{DeltaBeta}
\end{multline}
with $\Delta\beta^{(0)}(\omega_s, \omega_i)$ the phase-mismatch of a single waveguide.
When tuning the CW pump wavelength to the single-waveguide phase-matching resonance, and spectrally filtering the emitted state around degeneracy, we have $\Delta\beta^{(0)}=0$. Neglecting the possible small polarization dependence of the coupling constant, we further have $C_{s}=C_{i} \equiv C$ and thus $\Delta\beta = -2  (\cos k_s + \cos k_i ) C$. The wavefunction now only depends on the transverse wavevector and reads 
\begin{equation}
	\widetilde{\Psi}(k_s,k_i) = \gamma L \, \widetilde{A}_p(k_s+k_i) \, \sinc \left( f(k_s,k_i) \right)  e^{i f(k_s,k_i)}
	\label{Psi_k}
\end{equation}
with $ f(k_s,k_i) \!= \! (\cos k_s + \cos k_i ) C L$.
Decomposing the spatial Fourier transform of the pump beam as $\widetilde{A}_p(k_s+k_i) = \sum_n A_n e^{-i (k_s+k_i) n}$, where $A_n$ is the amplitude of the pump beam in waveguide $n$, the real-space wavefunction $\Psi (n_{s},n_{i}) $ can be obtained by Fourier transforming Eq.~\eqref{Psi_k}.
Using the integral representation of the cardinal sine, $\operatorname{sinc}(x)=\frac{1}{2} \int_{-1}^{1} e^{i x t} d t$, and performing a change of variables allows identifying integrals of the form $\frac{1}{2 \pi} \int_{-\pi}^{\pi}  e^{i 2C z\cos(k)} e^{i kn} \, dk = i^n J_n(2Cz)$. This leads to the following expression for the biphoton wavefunction in real space:
\begin{equation}
	\Psi (n_{s}, n_{i}) \! = \! \gamma \sum_n A_n \, i^{n_s+n_i-2n} \!\int_{0}^{L}  \!\! J_{n_s-n}(2Cz) J_{n_i-n}(2Cz) \, dz
\end{equation}
This expression establishes a direct connection between the quantum state produced by a linear and a nonlinear waveguide array, respectively \cite{SM}. Indeed, we can rewrite it as
\begin{equation}
	\Psi (n_{s},n_{i}) = \gamma \sum_n A_n \!\int_{0}^{L} \!\! \phi_n(n_{s},n_{i},z)  \,  dz 
	\label{Active}
\end{equation}
where
\begin{equation}
	\phi_n(n_{s},n_{i},z) = i^{n_{s} + n_{i}-2n}\ J_{n_{s}-n}(2Cz)\ J_{n_{i}-n}(2Cz) 
	\label{Passive}
\end{equation}
is the quantum state that would be generated, after a propagation distance $z$, by a linear array with two photons initially injected in waveguide $n$ (Eq.~\eqref{phi0} being an example of such state when $n=0$) \cite{Raymond24}.

\begin{figure*}[t]
	\centering
	\includegraphics[width=14cm]{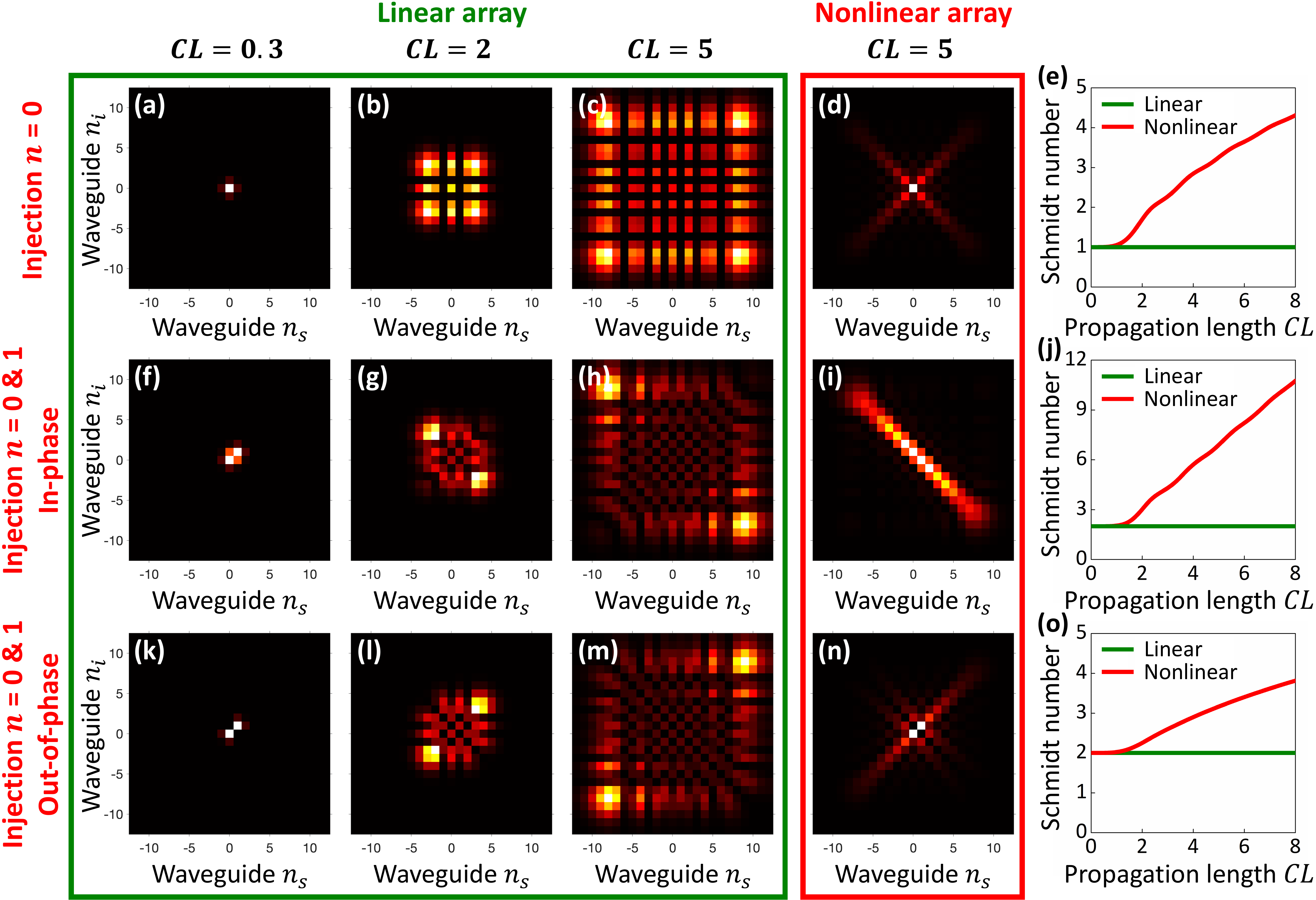}
	\caption{
		Comparison of quantum walks in linear and nonlinear waveguide arrays.
		First line (a-e): Simulated output correlation matrices when injecting two indistinguishable photons in the central waveguide ($n=0$) of a linear waveguide array, for increasing propagation lengths (a) $CL=0.3$, (b) $CL=2$ and (c) $CL=5$, compared to (d) the output of a nonlinear waveguide array pumped in the central waveguide; (e) calculated Schmidt number as a function of $CL$ for a linear (green line) and nonlinear (red line) array.
		Second line (f-j): same quantities, when injecting a path-entangled state $ (\ket{00} + \ket{11} )/\sqrt{2}$ in a linear array compared to pumping in-phase waveguides $n=0$ and $1$ of a nonlinear array.
		Third line (k-o): same quantities, when injecting a path-entangled state $ (\ket{00} -\ket{11} )/\sqrt{2}$ in a linear array compared to pumping out-of-phase waveguides $n=0$ and $1$ of a nonlinear array. All calculations are made using the analytical expressions of the quantum states provided in Eqs. \eqref{Active} and \eqref{Passive}.
	}
	\label{Fig_Passive}
\end{figure*}

Now coming back to the situation of a nonlinear waveguide array pumped in the central waveguide, Eq.~\eqref{Active} with $A_n=\delta_{0,n}$ yields the output quantum state
\begin{equation}
	\Psi_{n = 0}\left( n_{s},n_{i} \right) = \gamma \int_{0}^{L} \! \phi_{n = 0}(n_{s},n_{i},z)  \,  dz 
	\label{Psi0}
\end{equation}
The corresponding correlation map $\left| \Psi_{n = 0} \right|^{2}$ is shown in Fig.~\ref{Fig_Passive}d for $CL=5$.
As indicated by Eq.~\eqref{Psi0}, this state results from the superposition of states $\phi_{n=0}$ (Eq.~\eqref{phi0}) corresponding to propagation in linear arrays of lengths continuously ranging from $0$ to $L$.
We observe that, starting from the square pattern of these states (Fig.~\ref{Fig_Passive}a-b-c), their summation over the propagation distance generates a distinct interference pattern (Fig.~\ref{Fig_Passive}d), where primarily only the terms along the diagonal and antidiagonal remain.
This can be understood from the fact that these points can be reached by paths where the two photons of each pair make the exact same number of jumps (either in the same, or opposite directions), and thus accumulate the same propagation phase. 
Thus, on these points the biphoton amplitudes constructively interfere throughout the entire summation process over $z$, while at other points of the matrix the interference is destructive and averages out the biphoton amplitude.
This intuitive reasoning is confirmed by Eq.~\eqref{Psi0}, which shows that diagonal and antidiagonal points of the correlation matrix verify $ \vert\Psi_{n=0} (n,n)\vert = \vert\Psi_{n=0}(n, - n) \vert \propto \int_{0}^{L}{dz\ {{(J}_{n}(2Cz))}^{2}}$.
The integrand is positive, indicating that the states $ \phi_{n=0} $ interfere constructively along $ z $ on these points of the correlation matrix.
The resulting quantum state (Fig.~\ref{Fig_Passive}d) is entangled and as $CL$ increases, correlations spread across a greater number of waveguides, resulting in a continuous increase of the Schmidt number, as shown in Fig.~\ref{Fig_Passive}e (red line). 
This strongly differs from the case of the linear array (states $\phi_{n=0}$), where the Schmidt number remains constant at the value of the injected state ($K=1$, green line), as previously noted.

The emergence of this characteristic pattern—marked by dominant diagonal and antidiagonal elements in the correlation matrix—requires a minimum propagation length. This onset can be approximately identified by the condition 
$ \lvert \Psi_{n=0}(1,1) \rvert = \lvert \Psi_{n=0}(1,-1) \rvert > \lvert \Psi_{n=0}(0,1) \rvert $,
which signifies that the first diagonal element surpasses the first off-diagonal one. This condition is fulfilled when \( CL \gtrsim 1.2 \).
We note that this stabilization of the nonlinear-array state occurs at a higher value of $CL$ than for the corresponding linear-array state ($CL \simeq 0.72$), which is expected since the former arises from a summation of the latter: essentially, the correlation pattern of the nonlinear-array state is stabilized when the dominant off-diagonal terms corresponding to states $\phi_{n=0}$ with $CL<\!0.72$ are compensated by the dominant diagonal/antidiagonal terms of states $\phi_{n=0}$ with $CL>\!0.72$.

\begin{figure*}[t]
	\centering
	\includegraphics[width=14cm]{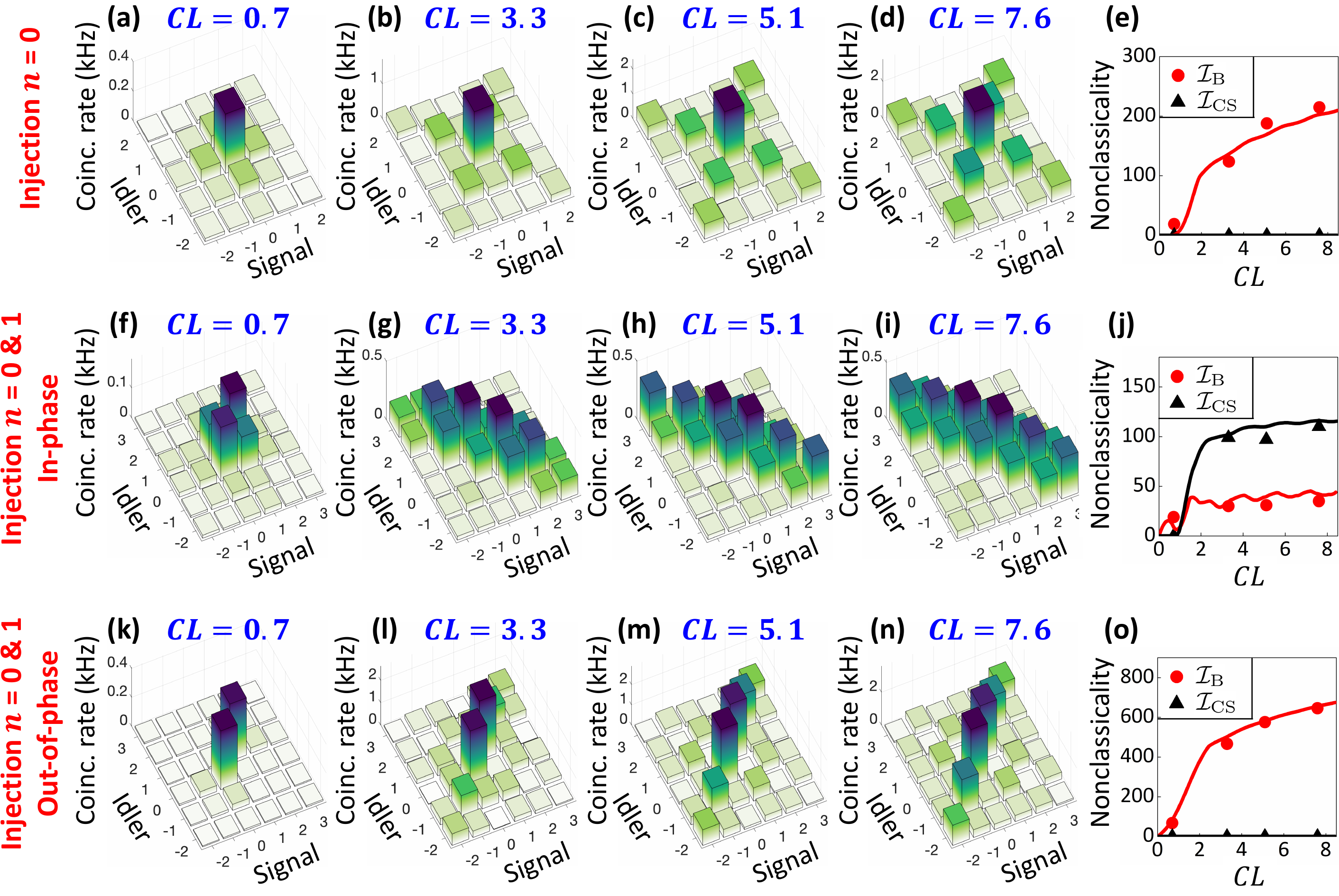}
	\caption{
		First line (a-e): Measured correlation matrices when pumping the central waveguide $n=0$, for (a) $CL=0.7$, (b) $CL=3.3$, (c) $CL=5.1$ and (d) $CL=7.6$, where $C=(C_{\rm TE}+C_{\rm TM})/2$ is the polarization-averaged coupling constant, and (e) experimental non-classicalities $\mathcal{I}_{\rm{B}} = \sum I_{\rm{B}}(n_s,n_i)$ (red dots) and $\mathcal{I}_{\rm{CS}}  = \sum I_{\rm{CS}}(n_s,n_i)$ (black triangles), as a function of $CL$, compared to theory (plain lines).
		Second line (f-j): same quantities, when pumping waveguides $n=0$ and $1$ in-phase.
		Third line (k-o): same quantities, when pumping waveguides $n=0$ and $1$ in phase opposition.
		In all cases the pump laser wavelength is tuned to the single-waveguide phase-matching resonance ($\lambda_{p}=783.45$ nm) and the total injected pump power is 1 mW. 
	}
	\label{Fig_Active}
\end{figure*}

\subsection{Experimental results}

To test these predictions and monitor the evolution of the correlation matrix as a function of $CL$, we performed experiments on a series of semiconductor AlGaAs nonlinear waveguide arrays fabricated with varying geometrical parameters, allowing to span $CL$ by a decade, from $0.7$ to $7.6$ \cite{SM}.
The different arrays are patterned by electron-beam lithography with a high-resolution HSQ resist followed by inductively coupled plasma (ICP) etching, starting from an epitaxial structure consisting of a 6-period Al$_{0.8}$Ga$_{0.2}$As/Al$_{0.25}$Ga$_{0.75}$As lower Bragg reflector, a 350 nm Al$_{0.45}$Ga$_{0.55}$As core and a 4-period Al$_{0.8}$Ga$_{0.2}$As/Al$_{0.25}$Ga$_{0.75}$As upper Bragg reflector.
The two Bragg reflectors simultaneously create a photonic bandgap for vertical confinement of the pump beam and act as total internal reflection claddings for the SPDC photons produced in the telecom range. As a result, the pump and SPDC modes exhibit distinct dispersion curves, enabling the generation of photon pairs via modal phase-matching \cite{Helmy11,Horn12,Francesconi23,Baboux23}.

The seven central waveguides of each array are linked to S-bent injection and collection waveguides, enabling the selective injection and retrieval of light in each waveguide (Fig.~\ref{Fig_Setup}c). These injection and collection waveguides feature a larger width, which shifts their nonlinear resonance wavelength, ensuring that the injected pump laser generates photon pairs exclusively within the central array region.
As sketched in Fig.~\ref{Fig_Setup}d, the experiments are performed by injecting a TE-polarized CW pump laser into the desired input waveguides of the nonlinear waveguide array to generate orthogonally polarized signal and idler photons by type-2 SPDC. 
The SPDC photons are collected through a lensed fiber array connected to superconducting nanowire single-photon detectors, after spectral filtering on a 16 nm bandwidth to select nearly frequency-degenerate photons. Two-photon coincidence events, recorded by a time tagger, provide the spatial correlation matrix $\Gamma_{n_s,n_i}=\vert \Psi (n_{s}, n_{i}) \vert^2$.

Figure~\ref{Fig_Active} displays the measured correlation matrices for (a) $ CL \! = \! 0.7 $, (b) $ CL \! = \! 3.3 $, (c) $CL \! = \! 5.1 $, and (d) $ CL \! = \! 7.6 $, where $C$ is the coupling constant averaged over both possible polarizations for the signal and idler photons (TE and TM), $C=(C_{\rm TE}+C_{\rm TM})/2$. The typical difference of coupling constant between both polarizations is $\sim 10 \%$, e.g. for (c) we have $C_{\rm TE}=2.7$ mm$^{-1}$ and $C_{\rm TM}=2.4$ mm$^{-1}$ (and $L=2$ mm).
In agreement with the theoretical analysis above, for $ CL \!< \!1.2 $, a transient structure with dominant off-diagonal terms is observed (Fig.~\ref{Fig_Active}a).  At higher $CL$, a stabilized regime emerges with dominant diagonal/antidiagonal terms, and correlations spread over more waveguides as \( CL \) increases (Figs.~\ref{Fig_Active}b to d).
We checked with full numerical simulations that for the investigated values of $CL$, the small polarization-dependence of the coupling constant has a negligible effect on the correlation matrix; specific effects, such as a fork-like splitting of the diagonal/antidiagonal lines of the correlation matrix (where each branch corresponds to one photon traveling at the "high" transverse velocity and the other photon at the "low" transverse velocity) would only appear at higher values of $CL$ \cite{Raymond24}. The generation rate of photon pairs at the chip output is of the order $10^5$  Hz/mW of internal pump power, corresponding to a brightness of the order of $10 $  kHz/mW/nm.

To quantify the non-classicality of the measured spatial correlations, two distinct criteria can be used. 
The first one, established by Bromberg \textit{et al.}~\cite{Bromberg09,Peruzzo10}, states that when two incoherent classical beams are coupled to two distinct input waveguides, the resulting intensity correlations $\Gamma_{n_s,n_i}=\vert \Psi (n_{s}, n_{i}) \vert^2$ always satisfy the following inequality, relating the off-diagonal correlations (\( \Gamma_{n_s, n_i} \) with \( n_s \neq n_i \)) to the corresponding diagonal correlations (\( \Gamma_{n_s, n_s} \) and \( \Gamma_{n_i, n_i} \)), through $\Gamma_{n_s,n_i} \!  >  \! \frac{2}{3}\sqrt{\Gamma_{n_s,n_s}\Gamma_{n_i,n_i}}$.
Based on this inequality, a non-classicality matrix $I_{\rm{B}}$ can be defined as
\begin{equation}
	I_{\rm{B}}(n_s,n_i) = {\rm{max}}\left(\frac{2}{3}\sqrt{\Gamma_{n_s,n_s}\Gamma_{n_i,n_i}}-\Gamma_{n_s,n_i}, \, 0\right)
	\label{Bromberg}
\end{equation}
On the other hand, the Cauchy-Schwarz (CS) inequality establishes that a classical state always obeys $\Gamma_{n_s,n_i} \!\! < \!\! \sqrt{\Gamma_{n_s,n_s}\Gamma_{n_i,n_i}}$. From this criterion, another non-classicality indicator $I_{\rm{CS}}$ can be evaluated as
\begin{equation}
	I_{\rm{CS}}(n_s,n_i) = {\rm{max}}\left(\Gamma_{n_s,n_i}-\sqrt{\Gamma_{n_s,n_s}\Gamma_{n_i,n_i}},\, 0\right)
	\label{Cauchy}
\end{equation}
These two indicators probe different aspects of spatial entanglement. For instance, the first indicator, \( I_{\rm{B}} \), is inherently insensitive to purely antidiagonal correlations, whereas the second indicator, \( I_{\rm{CS}} \), is particularly sensitive to them, so that both criteria are complementary.
We plot in Fig.~\ref{Fig_Active}e the total non-classicality $\mathcal{I_{\rm{B}}} \!=\!\sum_{n_s,n_i} \! I_{\rm{B}}(n_s,n_i)$ (red dots) and $\mathcal{I_{\rm{CS}}}\!=\!\sum_{n_s,n_i} \! I_{\rm{CS}}(n_s,n_i)$ (black triangles), normalized to the experimental error, for the various values of $C\!L$. For this quantum state, the Cauchy-Schwarz inequality is not violated. However, the inequality of Eq.~\eqref{Bromberg} is violated, with the degree of violation increasing progressively with $C\!L$. These experimental results are in good agreement with the theoretical predictions, shown with plain lines.

\section{Multiple-waveguide input}

Having clarified the dynamics of single-waveguide injection in both linear and nonlinear waveguide arrays,
we now turn to exploring multi-waveguide excitation as an efficient tool to engineer the output quantum state.

\subsection{Linear waveguide array}
\label{Sec_interference}

For this, we first consider the linear waveguide array case. Injecting a path-entangled state \mbox{$\left(\ket{00} \pm \ket{11}\right) \!/\! \sqrt{2}$} into neighboring waveguides ($n=0$ and $1$) leads, respectively, to the following output state, which we denote as $\phi_{n = 0 \pm 1}$:
\begin{equation}
	\begin{split}
		\phi_{n = 0 \pm 1}\left( n_{s},n_{i}, L \right) =
		i^{n_s+n_i} \big( J_{n_s} (2CL) J_{n_i} (2CL) \\
		\mp J_{n_s-1} (2CL) J_{n_i-1} (2CL) \big)
	\end{split}
	\label{phi01}
\end{equation}
The corresponding calculated correlation maps are shown in the second and third lines of Fig.~\ref{Fig_Passive}  (panels f-g-h and k-l-m, respectively), for increasing values of the normalized propagation length $CL$.
Compared to the separable state $\phi_{n = 0}$ discussed previously (panels a-b-c), we observe here a suppression of the diagonal (resp.\ antidiagonal) ballistic lobes, which leads to an entangled state with spatial antibunching (resp.\ bunching) character.

\begin{figure*}[t]
	\centering
	\includegraphics[width=0.52\textwidth]{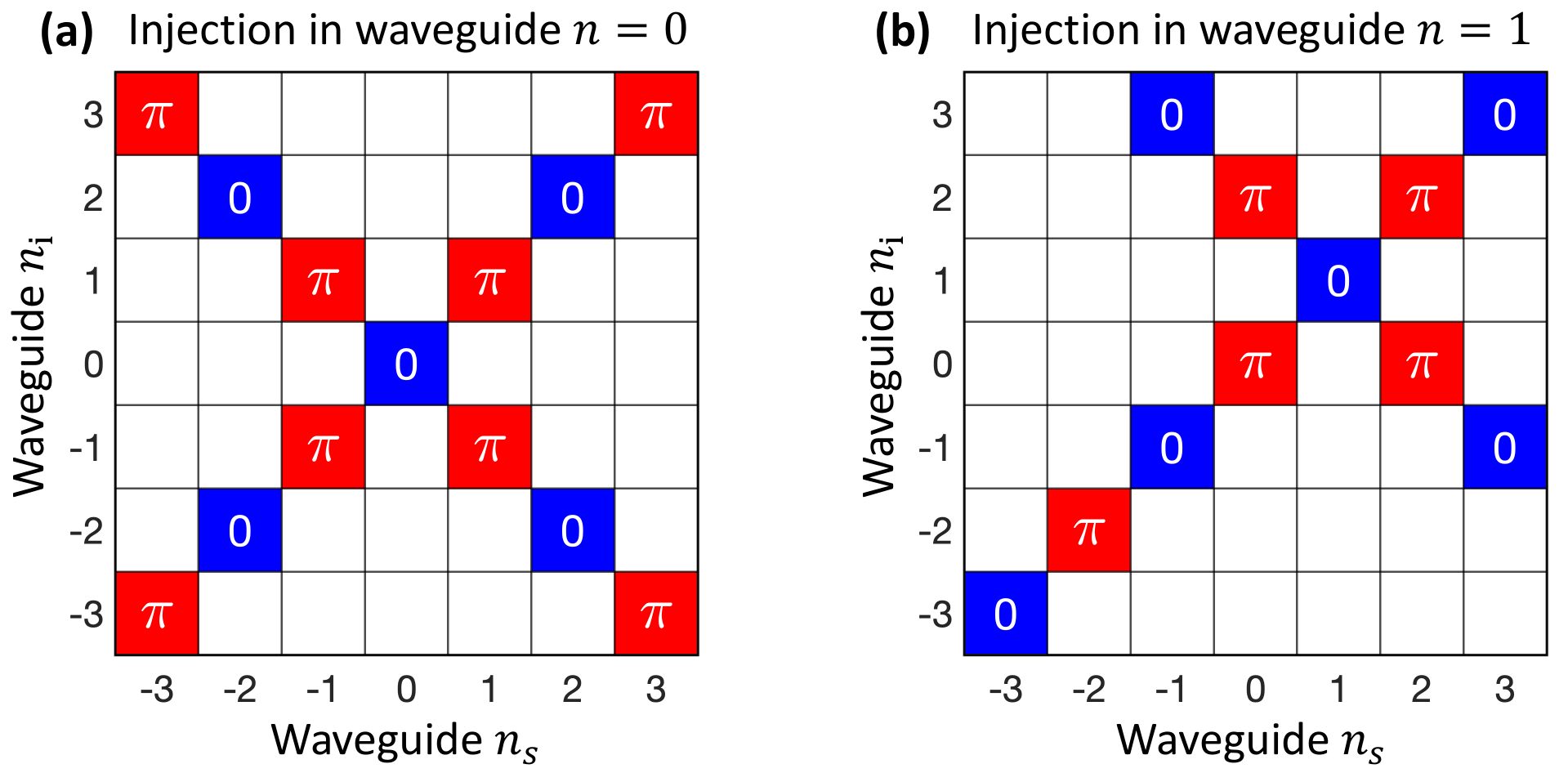}
	\caption{
		(a) Schematic representation of the phase structure of the quantum states $\phi_{n = 0}$ and $\Psi_{n = 0}$ along the diagonal and antidiagonal lines, and (b) corresponding schematic for the quantum states $\phi_{n = 1}$ and $\Psi_{n = 1}$.
	}
	\label{Fig_PhaseMatrix}
\end{figure*}

This behavior can be understood from a simple intuitive argument, in terms of constructive or destructive interference between the states $\phi_{n = 0}$ and $\phi_{n = 1}$ corresponding to injection of both photons in waveguide $0$ or $1$, respectively, since $\phi_{n = 0 \pm 1}=\phi_{n = 0} \pm \phi_{n = 1}$.
Let us first consider the $\phi_{n = 0}$ state. Coupled-mode theory predicts that a single photon acquires a phase shift of \( \pi/2 \) upon tunneling to an adjacent waveguide. As a result, starting from its injection in $n=0$, when the two-photon state moves one step along  the diagonal or antidiagonal of the correlation matrix—entailing two such single-photon tunnelings, whether on the same or opposite sides—it accumulates a total phase of \( \pi \).
Hence, the phase of the wavefunction switches by $\pi$ between every two successive points along the diagonal and antidiagonal, as sketched in Fig.~\ref{Fig_PhaseMatrix}a.
The phase structure of the $\phi_{n = 1}$ state is the same, but it is centered on the ($1,1$) point of the correlation map,  as sketched in Fig.~\ref{Fig_PhaseMatrix}b.
Therefore, the states $\phi_{n = 0}$ and $\phi_{n = 1}$ exhibit opposite phase along the diagonal line [points ($n,n$)], but same phase on the antidiagonal passing through $0$ and $1$ [points ($n,1-n$)].
This reasoning allows us to understand that for the output state \( \phi_{n = 0 + 1} \), corresponding to the injection of the path-entangled state \( (\ket{00} + \ket{11})/\sqrt{2} \), destructive interference occurs along the diagonal and constructive interference along the antidiagonal, resulting in a spatially antibunched state for sufficiently large $CL$.
Note that this behavior is reminiscent of the time-reversed Hong-Ou-Mandel effect on a beamsplitter \cite{Bromberg09,Fabre22}, where the injection of a bunched input results in an antibunched output.
Now turning to the state \( \phi_{n = 0 - 1} \), resulting from the injection of \( (\ket{00} - \ket{11})/\sqrt{2} \) at the array entrance, the reasoning is exactly reversed: the interference pattern is inverted, leading to a spatially bunched output state (Fig.~\ref{Fig_Passive}l,m).

Let us now analyze the stability of the correlation pattern with the propagation length.
For the state $\phi_{n = 0 + 1}$, the antibunching pattern stabilizes when the antidiagonal (passing between 0 and 1) becomes larger than the diagonal, i.e.\ when $\vert \phi_{n = 0 + 1}(1,0) \vert \geq \vert \phi_{n = 0 + 1}(0,0) \vert = \vert\phi_{n = 0 + 1}(1,1) \vert$, which occurs for $CL \gtrsim0.38$. This transition can be seen between the correlation maps of Figs.~\ref{Fig_Passive}f and g.
For the state $\phi_{n = 0 - 1}$, interestingly, all antidiagonal points are exactly zero:
\begin{equation}
	\begin{aligned} 
		\phi_{n=0-1} (n,1-n) =	i \,  \big(  & J_{n} (2Cz) J_{1-n} (2Cz) \\ 
		+ & J_{n-1} (2Cz) J_{-n} (2Cz) \big) = 0
	\end{aligned} 
\end{equation}
due to the Bessel function property $J_{-k}=(-1)^{k}J_{k}$. Hence, the bunching pattern of the $\phi_{n=0-1}$ state is established from arbitrarily small $CL$ (see Fig.~\ref{Fig_Passive}k).

As seen in Fig.~\ref{Fig_Passive}, moving from the left to the right columns, once the stable correlation pattern is established (at zero or non-zero $CL$, depending on the case), it essentially expands linearly with $CL$ but its shape remains qualitatively unchanged, i.e.\ either bunched or antibunched depending on the input state. However, this transverse spreading of the two-photon state does not increase the Schmidt number, which remains fixed to its initial value $K=2$ corresponding to the input state,  as shown in Figs.~\ref{Fig_Passive}j and o (green line): correlations only "dilute" over a larger number of waveguides as $CL$ increases.

We note that this stability of the correlation pattern, after a minimum propagation length, is in strong contrast to the case of two coupled waveguides, where e.g.\ for an input state \( (\ket{00} - \ket{11})/\sqrt{2} \) bunching and antibunching would alternate periodically along the propagation length. 
Indeed, a system of two coupled waveguides is equivalent to a beamsplitter with a reflectivity $R(L)=\sin^2(CL)$ that is periodic with the propagation length $L$, leading to the periodic appearance and disappearance of a given quantum state at the output. Such periodicity does not hold anymore in the case of a waveguide array: initially localized photons continuously and irreversibly leak towards outer waveguides (instead of bouncing back and forth due to boundary effects as in a two-waveguide system), leading, after a stabilization regime, to a progressive dilatation of the correlation profile while retaining the same overall shape.

\subsection{Nonlinear waveguide array}

We now turn to the nonlinear array case. The situation analogous to the one just studied consists in injecting the CW pump laser beam in two neighboring waveguides, either in phase or in phase opposition.
We denote the corresponding states as $\Psi_{n=0\pm1}$. Using Eq.~\eqref{Active} with an input pump profile $A_n=\delta_{0,n} \pm \delta_{1,n}$, we obtain
\begin{equation}
	\Psi_{n=0\pm1} \left(n_{s}, n_{i}\right)=\gamma \, i^{n_s+n_i}  \int_{0}^{L}  \phi_{n = 0 \pm 1} (n_{s},n_{i},z)\, dz
\end{equation}
This state thus results from the superposition of states of Eq.~\eqref{phi01} corresponding to propagation in linear arrays of lengths ranging between $0$ and $L$.
The corresponding correlation maps $\left| \Psi_{n=0 \pm 1} \right|^{2}$ are shown in Figs.~\ref{Fig_Passive}i and n, respectively, for $CL=5$.
Similarly to the single-waveguide pumping case (first line of Fig.~\ref{Fig_Passive}), we observe that the superposition of the quantum walks
started at all possible longitudinal positions produces a sharp interference pattern, which conserves, while strongly reinforcing the bunching or antibunching structure of the underlying linear-array states. 
Let us note that the generation of entanglement in nonlinear waveguide arrays originates from the interference of biphoton states (virtually) created at different transverse and longitudinal positions, and thus only requires the indistinguishability of these interfering biphotons. As a result, the effect is independent of the polarization of the individual photons composing the pairs, and type-2 SPDC, which produces cross-polarized photons, can be employed as effectively as type-0 or type-1 SPDC \cite{Raymond24}.

We now analyze each situation in more details.
When the two neighboring waveguides $n=0$ and $1$ are pumped in-phase, the diagonal terms read
\begin{equation}
	\Psi_{n=0+1} (n,n)=\gamma  (-1)^{n} \int_{0}^{L}  \big ( J_{n}^2(2Cz)-J_{n-1}^2(2Cz) \big ) \, dz
	\label{eqPsi01}
\end{equation}
while on the antidiagonal passing through the pumped waveguides, i.e.\ on the points $(n,1-n)$, by using the property $J_{-k}=(-1)^{k} \, J_{k}$ we obtain
\begin{equation}
	\Psi_{n=0+1} (n,1-n) = \gamma \, i \int_{0}^{L}  \! 2 J_{n}  (2Cz) J_{1-n} (2Cz) \, dz
\end{equation}
The latter terms are equal to twice the over-antidiagonal terms $\Psi_{n=0} (n,1-n)$ of the state corresponding to single-waveguide pumping. This is expected since, thanks to the linearity of Eq.~\eqref{Active} in $A_n$, the state $\Psi_{n=0+1} $ can be seen as resulting from the interference between the biphoton states $\Psi_{n=0}$ and $\Psi_{n=1}$ originating from each of the pumped waveguides. Due to the same argument detailed above regarding the phase structure of the two-photon state (related to the $\pi/2$ tunneling phase of single photons), these two states have same phase on the antidiagonal points $(n,1-n)$ and thus undergo constructive interference. Conversely, the destructive interference effect occurring along the diagonal is well apparent in the minus sign of the integrand of Eq.~\eqref{eqPsi01}.

The correlation pattern of the state $\Psi_{n=0+1} $, with dominant antidiagonal terms, is stabilized essentially when $\vert \Psi_{n = 0 + 1}(1,0) \vert \geq \vert \Psi_{n = 0 + 1}(0,0) \vert = \vert\Psi_{n = 0 + 1}(1,1) \vert$, which occurs for $CL \gtrsim 0.85$ (i.e. again later than for the corresponding linear-array state, for the same reason as detailed when discussing the $\Psi_{n=0} $ state). 
The calculated Schmidt number, shown in Fig.~\ref{Fig_Passive}j (red line), increases monotonically with the propagation length, as the antidiagonal correlations spread wider.
We note that the generation of such high-dimensional entangled states in a linear waveguide array would require the precise preparation of an input quantum state with the same modal dimensionality $K$ (requiring phase-controlled photon injection in at least $K$ waveguides). In contrast, nonlinear arrays exploit the constructive interplay between quantum walks and SPDC generation, enabling to control the output quantum state using only classical resources, by adjusting the spatial distribution of the input pump laser.

We now turn to the situation where the two neighboring waveguides $n=0$ and $1$ are pumped out-of-phase. Here, the diagonal terms read
\begin{equation}
	\Psi_{n=0-1} (n,n)=\gamma \, (-1)^{n} \int_{0}^{L}  \big ( J_{n}^2(2Cz) + J_{n-1}^2(2Cz) \big ) \, dz
\end{equation}
where the positive sign in the integrand here reflects constructive interference,
while the antidiagonal terms cancel out,
\begin{equation}
	\Psi_{n=0-1} (n,1-n) = 0
\end{equation}
since on these points the biphoton states $\Psi_{n=0}$ and $\Psi_{n=1}$ generated in the two pumped waveguides interfere with the exact same intensity but opposite phase, resulting in total destructive interference.
Hence, for the state $\Psi_{n=0-1}$ the bunching pattern (with dominant diagonal terms) is established from arbitrary small $CL$, and $K$ again increases monotonically with $CL$, as diagonal correlations spread farther, as seen in Fig.~\ref{Fig_Passive}o (red line).

We note that the asymmetry between bunched and antibunched states, regarding the stability of their correlation pattern as a function of the propagation length, can be related to the fact that the diagonal and antidiagonal points of the correlation matrix have a distinct status (as soon as several waveguides are pumped) due to the localized character of SPDC generation, which favors diagonal correlations at short propagation length. 
The bunched state $\Psi_{n=0-1}$ is thus not simply the "symmetric" of the antibunched state $\Psi_{n=0+1}$. 
This contrasts with the linear-array states $\phi_{n=0+1}$ and $\phi_{n=0-1}$, which appear as essentially symmetric of each other at sufficiently large propagation length (compare e.g. Figs.~\ref{Fig_Passive}h and m), with bunching playing a similar role as antibunching.
In other words, the local generation of photon pairs in nonlinear waveguide arrays breaks the symmetry between the bunching and antibunching features of the underlying linear-array states.

\subsection{Experimental results}

We now proceed to testing these theoretical predictions with experiments. 
We use the same nonlinear waveguide arrays as in Figs.~\ref{Fig_Active}a-d (single-waveguide pumping) but we here coherently pump two neighboring waveguides ($n=0$ and $1$) by using a free-space beamsplitter to split the injection beam into two parts, with the relative phase controlled by a wedged plate inserted in one arm.
This enables the injection of two laser beams with equal intensity (0.5 mW) and controlled phase relationship between them.
Figures~\ref{Fig_Active}f-g-h-i show the correlation matrices measured when pumping two neighboring waveguides in-phase (state $\Psi_{n=0+1}$) for increasing values of $CL$, from $CL=0.7$ to $CL=7.6$.
In good agreement with the theory described above, for \( CL < 0.85 \) (Fig.~\ref{Fig_Active}f), we still observe a transient pattern with dominant diagonal terms. At longer propagation length (Figs.~\ref{Fig_Active}g to i), the biphoton state reaches a stabilized regime with dominant antidiagonal terms, and the correlations expand further as \( CL \) increases. 
Figure~\ref{Fig_Active}j shows the corresponding experimental non-classicality indicators $\mathcal{I_{\rm{B}}} $ (red dots) and $\mathcal{I_{\rm{CS}}}$ (black triangles), as determined from Eq.~\eqref{Bromberg} and Eq.~\eqref{Cauchy} respectively, compared to the theoretical predictions (plain lines).
The first indicator here exhibits significantly lower values compared to those obtained for the state $\Psi_{n=0}$  (Fig.~\ref{Fig_Active}e, red dots) since, as mentioned earlier, this indicator is by construction largely insensitive to antidiagonal correlations. By contrast, the Cauchy-Schwarz indicator $\mathcal{I_{\rm{CS}}}$, which is particularly sensitive to spatial antibunching, is here violated and takes higher values than $\mathcal{I_{\rm{B}}} $.

Figures~\ref{Fig_Active}k-l-m-n show the correlation matrices measured when pumping the same two waveguides but in phase opposition (state $\Psi_{n=0-1}$), for increasing values of $CL$.
In agreement with the theoretical predictions, and in contrast to the previous case, the correlation pattern is here established from very low propagation lengths (Fig.~\ref{Fig_Active}k) and persists as the propagation length increases, with correlations expanding wider along the diagonal (Figs.~\ref{Fig_Active}l to n).
The corresponding non-classicality indicators $\mathcal{I_{\rm{B}}} $ and $\mathcal{I_{\rm{CS}}}$ are shown in Fig.~\ref{Fig_Active}o.
Their behavior here is qualitatively similar to that of the quantum state $\Psi_{n=0}$ (Fig.~\ref{Fig_Active}e), with zero violation of the Cauchy-Schwarz inequality and on the contrary increasingly large violation of the inequality of Eq.~\eqref{Bromberg} as $CL$ increases.

\section{Inverse design of nonlinear waveguide arrays}

We have shown up to now, both theoretically and experimentally, how pumping two waveguides of a nonlinear waveguide array can be exploited to generate various types of path-entangled biphoton states. Generalizing this principle, a broader variety of quantum states could be achieved by simultaneously pumping more waveguides, with controlled intensity and phase relations between them. To implement this efficiently and scalably, on-chip beamsplitters and phase shifters, leveraging the strong electro-optic effect of AlGaAs \cite{Wang14}, could be integrated prior to the generation stage.

\begin{figure*}[t]
	\centering
	\includegraphics[width=13.46cm]{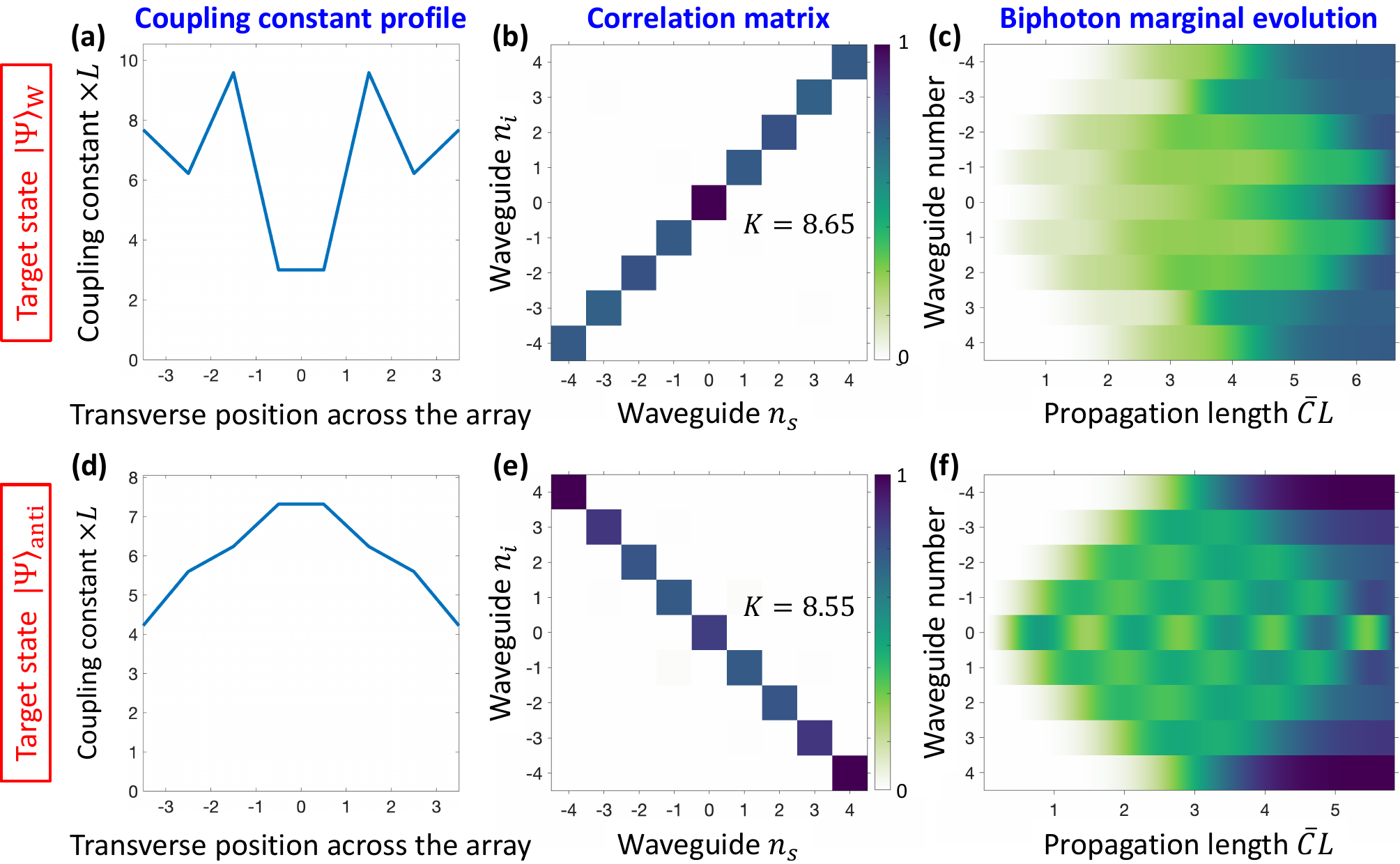}
	\caption{
		Examples of quantum state engineering of a biphoton W-state (top row) and a perfectly anticorrelated state (bottom row) in aperiodic nonlinear waveguide arrays with $N=9$ waveguides.
		(a) Transverse spatial profile of the coupling constants for the W-state, (b) output correlation matrix (with Schmidt number $K$), and (c) evolution of the marginal $I(n_s)=\sum_{n_i}\vert \Psi (n_{s}, n_{i}) \vert^2$ of the correlation matrix as a function of the normalized propagation length $\overline{C}L$, where $\overline{C}$ is the average of the coupling constants across the transverse direction. 
		(d–f)~Same quantities for the perfectly anticorrelated state.
		In the coupling profile plots (a) and (d), the value of the coupling constant between neighboring waveguides $n$ and $n+1$ is represented at the intermediate position $n + 1/2$.
	}
	\label{Fig_Engineering}
\end{figure*}

A complementary approach consists in engineering the couplings between waveguides to realize \textit{aperiodic} instead of periodic, homogeneously coupled arrays. This can be achieved either in a fixed manner, by setting the inter-waveguide distances during fabrication, or in a reconfigurable manner, by exploiting again the electro-optic effect of AlGaAs.
Combining both strategies, i.e.~simultaneously tailoring the pump profile and engineering the coupling constants, is expected to further expand the possibilities of quantum state engineering in nonlinear waveguide arrays.
To illustrate the potential of this approach, let us consider two target states corresponding to maximally entangled biphoton states: the state $\ket{\Psi}_{\rm W}$, exhibiting perfect diagonal correlations, and the state $\ket{\Psi}_{\rm anti}$, exhibiting perfect antidiagonal correlations. For an array of $N$ waveguides, these states are defined as:
\begin{equation}
	\ket{\Psi}_{\rm W} = \frac{1}{\sqrt{N}} \sum_{n=-\bar{n}}^{\bar{n}}   \! e^{i \varphi_n} \ket{n,n}_{s,i}
\end{equation}
\begin{equation}
	\ket{\Psi}_{\rm anti} = \frac{1}{\sqrt{N}} \sum_{n=-\bar{n}}^{\bar{n}}   \! e^{i \varphi_n} \ket{-n,n}_{s,i}
\end{equation}
where $\bar{n}=(N-1)/2$ and the kets indicate the output waveguide of signal and idler photons, respectively. The phases $\varphi_n$ can be arbitrarily chosen without affecting the maximally entangled character of the state (in the case of an array of identical waveguides, photon tunneling naturally favors $\varphi_n=(-1)^n$, as noted earlier).
The first state corresponds to both photons bunching together in the same waveguide, with equal probability between all possible waveguides --- a biphoton version of a single-photon W-state \cite{Grafe14}, where a photon is equally spread across all basis modes of the Hilbert space. The second state corresponds instead to both photons always exiting through opposite waveguides. While hints of such states can be observed in homogeneously coupled arrays (Fig.~\ref{Fig_Active}), we will now show through numerical simulations that a combined engineering of the pump and coupling profiles enables their near-perfect realization.

To keep experimental resources simple, we consider an array of $N=9$ waveguides, where the central three waveguides are pumped with complex amplitudes $A_{-1}$, $A_0$, and $A_1$. Due to the symmetry of the target states, we assume a symmetric pump profile, yielding two free parameters: the ratio $a=\vert A_{1}/A_{0} \vert=\vert A_{-1}/A_{0} \vert$ between the pumping strength in the central and first neighbor waveguides, and their phase relation $\varphi=\text{arg}(A_{1}/A_{0})=\text{arg}(A_{-1}/A_{0})$. The coupling constants are also assumed to be symmetric about the center waveguide.

We implemented a gradient descent algorithm to jointly optimize the pump and coupling profiles to maximize the similarity
$S= \big( \sum \sqrt{ \Gamma_{n_s,n_i}^{\smash{\raisebox{0.1ex}{\scriptsize\rm gen}}}  \Gamma_{n_s,n_i} ^{\rm tar}}  \big)^2 / \left(  \sum  \Gamma_{n_s,n_i}^{\smash{\raisebox{0.1ex}{\scriptsize\rm gen}}} \sum  \Gamma_{n_s,n_i} ^{\rm tar}   \right)$ between the generated ($\Gamma_{n_s,n_i}^{\smash{\raisebox{0.1ex}{\scriptsize\rm gen}}}$) and target ($\Gamma_{n_s,n_i} ^{\rm tar}$) intensity correlation matrices (a reliable proxy for fidelity since the phases $\varphi_n$ do not affect entanglement). The optimized parameters are listed in Table~\ref{Table_Engineering}, along with figures of merit for both the W and antidiagonal states.
Figure~\ref{Fig_Engineering} displays, for each case, the optimized spatial coupling profile (left), the output correlation matrix (center), and the evolution of the marginal biphoton intensity $I(n_s)=\sum_{n_i}\vert \Psi (n_{s}, n_{i}) \vert^2$ (right) as a function of the normalized propagation length $\overline{C}L$, where $\overline{C}$ is the average of the coupling constants across the transverse direction. For both states, we achieve a similarity above 98.5\% for the correlation matrix, corresponding to state fidelities above 98.5\% and Schmidt numbers above 8.55, close to the maximum value of 9 corresponding to the dimension of the considered Hilbert space.
Interestingly, we found that the output quantum states are appreciably robust to potential experimental errors: e.g.~for the biphoton W-state, a 10\% variation in the coupling constants and pump parameters (relative to their optimal values) still maintains the similarity and fidelity above 90\%, highlighting the practical relevance of this approach for a realistic experimental implementation. Note that we here considered (in agreement with previous Sections) an uncoupled pump beam, but if required, a possible small transverse propagation of the pump beam could be readily included in the inverse-design algorithm. We note also the insightful analogy between our situation—engineering aperiodic coupling profiles to realize particular \textit{spatially} entangled states—and the design of aperiodic poling patterns in dielectric crystals to target specific \textit{spectrally} entangled states \cite{Graffitti17,Graffitti20}.

\begin{table}[h]
	\caption{Parameters and figures of merit for quantum state engineering in an array of 9 nonlinear waveguides. Coupling coefficients $C_1$ to $C_4$ (in units of $L^{-1}$) are listed from the center outward. Pump parameters $a$ and $\varphi$ are defined in the text.}
	\centering
	\begin{tabular}{ccccccccc}
		\hline
		\hline
		Target state & $C_1$ & $C_2$ & $C_3$ & $C_4$ & $a$ & $\varphi$ & Similarity & $K$ \\
		\hline
		$\ket{\Psi}_{\rm W}$        & 3.00 & 9.58 & 6.22 & 7.68 & 2.40  & $\pi$ & 0.990 & 8.65 \\
		$\ket{\Psi}_{\rm anti}$     & 7.22 & 6.36 & 5.66 & 4.28 & 0.424 & 0    & 0.985 & 8.55 \\
		\hline
		\hline
	\end{tabular}
	\label{Table_Engineering}
\end{table}

For both investigated quantum states, the marginal intensity $I(n_s)=\sum_{n_i}\vert \Psi (n_{s}, n_{i}) \vert^2$ evolves in a non-trivial way (see Figs.~\ref{Fig_Engineering}c and \ref{Fig_Engineering}f), including edge-reflections of photons at half propagation length, indicating that state generation leverages both interference and finite-size effects. This emphasizes the relevance of a gradient-based inverse-design approach---complementary to powerful recently developed analytical or semi-analytical methods \cite{Luo19,Barral20,Belsley20,Barral21,He24,Costas25,Delgado25}--- to determine the optimal coupling and pumping parameters, given the complexity of intuition-based reasoning here.
It can be noticed, however, that the optimal pump phases ($\varphi=0$ and $\pi$, as seen in Table~\ref{Table_Engineering}) match those used for the homogeneously coupled arrays in Fig.~\ref{Fig_Active} (second and third lines): in-phase pumping favors antidiagonal correlations, while out-of-phase pumping enhances diagonal ones, consistently with the argument on constructive versus destructive interference presented in Section~\ref{Sec_interference}. However, Table~\ref{Table_Engineering} also shows that the optimal pump amplitude ratios $a$ differ significantly from 1, and in opposite directions for the two states, highlighting again the limitations of a purely intuitive approach.

Regarding the coupling profiles, they also exhibit markedly different characteristics: the antidiagonal state requires a smooth profile with monotonically decaying coupling constants on either side of the central waveguides (Fig.~\ref{Fig_Engineering}d), whereas the diagonal state necessitates a more complex, non-trivial variation (Fig.~\ref{Fig_Engineering}a).
It is interesting to compare the latter profile with the one considered for the generation of single-photon W-states in Ref.~\cite{Grafe14}, where a single photon is injected into the central waveguide of a linear (passive) waveguide array. 
The overall shape of the corresponding coupling profile, featuring minimal couplings at the center, shares some similarity with the one of Fig.~\ref{Fig_Engineering}a. However, using such a linear array with two photons injected into the central waveguide would result in a perfectly flat (uncorrelated) intensity matrix, corresponding to an equal probability for each photon to exit through any waveguide — a situation in stark contrast with the biphoton W-state, which is maximally entangled. In any case, the biphoton W-state, whose feasibility we demonstrate here, would be challenging to produce in any linear waveguide array, as it would require an input state with Schmidt number $K \approx N$, achievable only with a biphoton spatial profile spanning all $N$ waveguides, which is experimentally demanding.
In contrast, our approach involves a simpler three-waveguide pumping scheme with a laser. This again illustrates the effectiveness of nonlinear waveguide arrays for generating spatially entangled quantum states, thanks to the constructive interplay between quantum walks and SPDC generation that they allow.

\section{Conclusion and perspectives}

In summary, we conducted a systematic comparison of quantum walks in linear and nonlinear waveguide arrays, combining analytical and numerical approaches. These predictions were validated through experiments in AlGaAs semiconductor nonlinear waveguide arrays with varying parameters, enabling tunability of the coupling–propagation length product $CL$, and thus of the depth of the quantum walks, over one order of magnitude. We analyzed the evolution of the correlation patterns with $CL$  and examined the gradual onset of nonclassical behavior using two complementary criteria, each highlighting different features of path entanglement. Finally, we introduced an inverse-design strategy to tailor aperiodic waveguide lattices with spatially varying coupling constants, demonstrating that this approach can produce maximally entangled states with moderate experimental overhead.

As a general outlook, we emphasize that linear and nonlinear approaches, rather than being mutually exclusive, can be advantageously combined within the same monolithic chip. Indeed, by locally adjusting the waveguide width, one can shift the modal refractive index and thus tune the nonlinear phase-matching wavelength, bringing the parametric process in or out of resonance with the pump laser. This provides a simple means to switch photon-pair generation on or off, in a manner analogous to the sequencing of poled and unpoled regions in dielectric platforms such as lithium niobate \cite{Sansoni17,Chapman25}.
Such control could enable the design of integrated devices where a nonlinear section efficiently generates a maximally entangled state—such as the perfectly diagonal or antidiagonal state—with minimal experimental resources, followed by a well-tailored linear array section that transforms this state into other maximally entangled configurations, chosen at will by engineering the coupling profile or/and the on-site energies of this linear section. Such a hybrid approach would significantly expand the range of quantum states that can be produced by waveguide arrays while keeping a reasonably simple pump configuration (three pumped waveguides in this case).

Besides such quantum state engineering applications, combining nonlinear and linear sections on the same chip could also allow quantum state tomography. Indeed, discrete fractional Fourier transforms can be realized in waveguide arrays through engineered coupling profiles that mimic the $J_x$ operator algebra~\cite{Christandl05,Chen16_Feder,Weimann16}. Measuring the intensity correlation matrices in both real and momentum space (i.e., after the Fourier transform) would enable phase retrieval of the biphoton wavefunction using dedicated reconstruction algorithms \cite{Orlowski94,Dehghan24}.
Altogether, such integration of generation, manipulation, and characterization of quantum states of light highlights the strong potential of continuously coupled photonic systems to exploit the high-dimensional spatial degrees of freedom of photons within compact architectures.
In parallel, the ability to freely tune the coupling and propagation constants of the waveguides can enable the exploration of a wide range of physical phenomena—including Anderson localization \cite{DiGiuseppe13,Bai16}, topological protection \cite{BlancoRedondo18,Leykam15,Bergamasco19,Cheng19}, dynamic localization \cite{Tang22}, or synthetic dimensions \cite{Piccioli22}—in a controlled environment, thereby opening promising avenues for implementing quantum simulation tasks on-chip.

\section*{Acknowledgments}

We thank M.I.~Amanti for fruitful discussions and P.~Filloux for technical support. We acknowledge support from the Ville de Paris Emergence program (\textsc{LATTICE} project), Region Ile de France in the framework of the DIM QuanTiP (\textsc{Q-LAT} project), IdEx Université Paris Cité (ANR-18-IDEX-0001) and the French \textsc{RENATECH} network.

\end{document}